\documentclass[a4paper]{article}

\usepackage{INTERSPEECH2020}
\usepackage{listings}
\usepackage{multicol, multirow}
\usepackage[hyphens]{url}

\title{Surfboard: Audio Feature Extraction for Modern Machine Learning
}
\name{Raphael Lenain, Jack Weston, Abhishek Shivkumar, Emil Fristed}
\address{Novoic Ltd}
\email{\{raphael, jack, abhishek, emil\}@novoic.com}

\begin{document}

\maketitle
\begin{abstract}

We introduce Surfboard, an open-source Python library for extracting audio features with application to the medical domain. Surfboard is written with the aim of addressing pain points of existing libraries and facilitating joint use with modern machine learning frameworks. The package can be accessed both programmatically in Python and via its command line interface, allowing it to be easily integrated within machine learning workflows. It builds on state-of-the-art audio analysis packages and offers multiprocessing support for processing large workloads. We review similar frameworks and describe Surfboard's architecture, including the clinical motivation for its features. Using the mPower dataset, we illustrate Surfboard's application to a Parkinson's disease classification task, highlighting common pitfalls in existing research. The source code is opened up to the research community to facilitate future audio research in the clinical domain.

\end{abstract}

\section{Introduction} 

The diversity of applications of acoustic analysis is best demonstrated by the last 10 years of the INTERSPEECH Computational Paralinguistics Challenges\footnote{\url{http://www.compare.openaudio.eu/}}, encompassing emotion detection \cite{compare2009}, gender prediction \cite{compare2010}, speaker state trait prediction \cite{compare2011,compare2012} and detection of medical conditions \cite{compare2013,compare2015} to name a few. Acoustic speech changes have been identified in a multitude of motor disorders (e.g. Parkinson's disease) \cite{little2008suitability,tsanas2012novel,orozco2016automatic}, affective disorders (e.g. depression) \cite{low2020automated} and respiratory diseases (e.g. pneumonia) \cite{abeyratne2013cough,pramono2016cough}. These acoustic changes can often be detected using features extracted from speech and it is common for papers with a clinical application to define their own feature sets \cite{little2008suitability, tsanas2012novel, pramono2016cough}. However, the selection of features and how to extract them is inconsistent across the field, resulting in a need for harmonization.

Surfboard is a Python package for audio feature extraction, written with the aim of making a library better suited to fast prototyping and modern machine learning (ML) applications than what is offered today. Our work is most similar to OpenSMILE \cite{eyben2010openSMILE}, an audio feature extractor implemented in C++ that was first released in 2010 and had its latest release in 2016. OpenSMILE extracts `low-level descriptors' (LLDs) from audio signals and combines them with `functionals', functions that operate on time series data to extract time-independent features. Examples of LLDs include the mel-frequency cepstrum coefficients (MFCCs) and the loudness; statistics include mean and standard deviation over time.

OpenSMILE is computationally efficient but custom configuration is complex. There exists no simple Python interface for OpenSMILE, hindering use in conjunction with modern ML frameworks such as scikit-learn \cite{pedregosa2011scikit}, TensorFlow \cite{abadi2016tensorflow} and PyTorch \cite{paszke2019pytorch}. Praat \cite{boersma2018praat} is another popular audio feature extractor, first released in 1991. Praat is desktop-based software which since 2018 has been complemented by a Python wrapper, Parselmouth \cite{jadoul2018introducing}, and suits detailed analysis of small numbers of audio files. MATLAB \cite{MATLAB:2010} is frequently used by members of the audio community to extract features from speech and music, for example using Voicebox \cite{brookes1997voicebox}, MIRtoolbox \cite{lartillot2007MATLAB} or, more recently, Audio Toolbox. While these are reliable toolkits, working within a MATLAB environment adds an unnecessary constraint to audio feature extraction and inhibits usability. In designing Surfboard, we attempt to combine the best of all these approaches to suit multiple use cases, including large-scale deployment.


In this paper, we first describe Surfboard's architecture, interface, audio features and the clinical rationale behind the features. We compare these with features common to both OpenSMILE and Praat. Finally, we present a ML classifier trained on Surfboard features extracted from part of the mPower dataset \cite{bot2016mpower} to highlight issues with prior work in Parkinson's, and provide a list of reference values derived from the LibriSpeech dataset \cite{panayotov2015librispeech}. We release the Surfboard codebase\footnote{\url{https://github.com/novoic/surfboard}} to the research community under an open-source license, along with notebooks containing all the code used in this paper\footnote{\url{https://github.com/novoic/surfboard-IS2020}}.

\begin{table*}[h!]
\caption{Description of the Surfboard features, including implementation, reference values and clinical rationale. Surfboard v0.1 reference values based on a 40-hour subset of LibriSpeech \cite{panayotov2015librispeech} are provided. The right half of the table is adapted from \cite{boschi2017connected,chan2019hd,noffs2018speech,low2020automated} and summarizes clinical validation of recent review papers across indications. $\uparrow$ = feature increases compared with healthy controls; $\downarrow$ = feature decreases compared with healthy controls; $\updownarrow$ = feature can increase or decrease compared with healthy controls, depending on derived feature (e.g. which MFCC component). $\leftrightarrow$ is used to indicate that features have been applied for classification, but that how they change is unknown. For WhC, the symbol $\leftrightarrow$ has been used: none of the reviewed papers on respiratory conditions reported feature values but the inclusion of spectral features was indeed motivated by the respiratory literature. - = unknown. PD = Parkinson's disease; MND = motor neurone disease, synonymous with amyotrophic lateral sclerosis (ALS); MS = multiple sclerosis; HD = Huntington's disease; MDD = major depressive disorder; HpM = hypomania; Anx = anxiety; Szo = schizophrenia; PTSD = post-traumatic stress disorder; WhC = whooping cough, synonymous with pertussis. A dagger ($\dagger$) indicates that the feature is a time series, so the reference value shown is the mean over time. * The pitch period entropy (PPE) reference method \cite{ppe} sometimes produces large negative outliers and the crest factor method \cite{boyd1986multitone} large positive outliers, so we show here the more meaningful median and the median absolute deviation statistics in lieu of the mean and standard deviation; we also note that PPE was developed to assess sustained phonations rather than free speech. Novel, robust implementations of these features will be added in a future version of Surfboard.} 
\centering 
\resizebox{\textwidth}{!}
{
\begin{tabular}{|c| c c|| c c c c c c c c c c|}
\hline\hline 
Component & Impl. & LibriSpeech & PD & MND & MS & HD & MDD & HpM & Anx & Szo & PTSD & WhC \\ [0.5ex] 
\hline 

\textbf{Entire waveform representations} & & & & & & & & & & & & \\
\hline
MFCCs & LibROSA & - & - & $\updownarrow$\cite{tsanas2011nonlinear,tsanas2012novel,bocklet2013}&-& $\uparrow$\cite{rusz2014phonatory} & $\downarrow$\cite{low2020automated} & $\downarrow$\cite{low2020automated} & $\downarrow$\cite{low2020automated} &-&-& $\leftrightarrow$\cite{pramono2016cough} \\

Log mel spectrogram & LibROSA & - &-&-&-&-&-&-&-&-&-&-\\

Morlet continuous wavelet transform & SciPy \cite{virtanen2020scipy} & - &-&-&-&-&-&-&-&-&-&-\\

Bark spectrogram & Ours & - &-&-&-&-&-&-&-&-&-&-\\

Magnitude spectrum & LibROSA & - &-&-&-&-&-&-&-&-&-&-\\ \hline

\textbf{Chromas (music motivated)} & & & & & & & & & & & & \\
\hline
Chromagram with STFT & LibROSA & - &-&-&-&-&-&-&-&-&-&-\\

Chromagram with CQT & LibROSA & - &-&-&-&-&-&-&-&-&-&-\\

Chroma CENS & LibROSA & - &-&-&-&-&-&-&-&-&-&-\\ \hline

\textbf{Spectral features} & & & & & & & & & & & & \\
\hline
Spectral slope$\dagger$ & Ours & $(-1.10\pm0.412)\times10^{-3}$  &-&-&-&-&-&-&-&-&-&$\leftrightarrow$\cite{pramono2016cough} \\

Spectral flux$\dagger$ & Ours & $(15.2\pm5.64)\times10^{-3}$ &-&-&-&-&-&-&-&-&-&-\\

Spectral entropy$\dagger$ & Ours & $4.46\pm0.352$ &-&-&-&-&-&-&-&-&-&-\\

Spectral centroid$\dagger$ & Ours & $(1.70\pm0.401)\times 10^3$ Hz &-&-&-&-&-&-&-&-&-&-\\

Spectral spread$\dagger$ & Ours & $(1.50\pm0.178)\times10^3$ Hz &-&-&-&-&-&-&-&-&-&$\leftrightarrow$\cite{pramono2016cough} \\

Spectral skewness$\dagger$ & Ours & $(1.74\pm0.621)\times10^{-3}$&-&-&-&-&-&-&-&-&-& $\leftrightarrow$\cite{pramono2016cough}\\

Spectral kurtosis$\dagger$ & Ours & $-2.99\pm0.00443$ &-&-&-&-&-&-&-&-&-& $\leftrightarrow$\cite{pramono2016cough}\\

Spectral flatness$\dagger$ & LibROSA & $(1.86\pm15.4)\times10^{-3}$ &-&-&-&-&-&-&-&-&-& $\leftrightarrow$\cite{pramono2016cough}\\

Spectral rolloff$\dagger$ & LibROSA & $(3.13\pm0.677)\times10^3$ Hz &-&-&-&-&-&-&-&-&-& $\leftrightarrow$\cite{pramono2016cough}\\ \hline

\textbf{Classical speech features} & & & & & & & & & & & & \\
\hline
F0 contour$\dagger$ & pysptk & $149 \pm 35.6$ Hz & $\updownarrow$\cite{bocklet2013} & $\updownarrow$\cite{tomik2010dysarthria} & $\updownarrow$\cite{noffs2018speech} & $\downarrow$\cite{rusz2014phonatory} & $\downarrow$\cite{low2020automated} & $\uparrow$\cite{low2020automated} & $\uparrow$\cite{low2020automated} & $\uparrow$\cite{low2020automated} & - & - \\

F0 SD & pysptk & $26.5\pm10.7$ Hz & $\downarrow$\cite{boschi2017connected,skodda2011intonation,bocklet2013} & $\downarrow$\cite{tomik2010dysarthria} & $\downarrow$\cite{noffs2018speech} & $\uparrow$\cite{rusz2014phonatory,chan2019hd} & $\uparrow$\cite{low2020automated} & - & $\uparrow$\cite{low2020automated} & $\downarrow$\cite{low2020automated} & $\downarrow$\cite{low2020automated} & - \\

Intensity$\dagger$ & Ours & $(4.16 \pm 5.63) \times 10^{-3} $ & - & - & - & - & $\downarrow$\cite{low2020automated} & - & $\downarrow$\cite{low2020automated} & $\downarrow$\cite{low2020automated} & - & - \\

Intensity SD & Ours &  $(6.33 \pm 5.61) \times 10^{-3} $ & $\downarrow$\cite{boschi2017connected} & - & - & - & $\downarrow$\cite{low2020automated} & - & - & $\uparrow$\cite{low2020automated} & - & - \\

Sliding-window root mean square (energy)$\dagger$ & LibROSA & $0.0444\pm0.0201$ & $\updownarrow$\cite{boschi2017connected} & - & $\downarrow$\cite{noffs2018speech} & - & - & - & - & - & - & - \\

Log energy & Ours & $-25.0\pm3.22$ & $\uparrow$\cite{tsanas2011nonlinear} & - & - & - & - & - & - & - & - & - \\

Sliding-window log energy$\dagger$ & Ours & $-34.7\pm4.81$ & - & - & - & - & - & - & - & - & - & - \\

Zero-crossing rate & LibROSA & $0.0528\pm0.0183$ & - & - & - & - & - & - & - & - & - & $\leftrightarrow$\cite{pramono2016cough} \\

Sliding-window zero-crossing rate$\dagger$ & LibROSA & $0.0527\pm0.0182$ & - & - & - & - & - & - & - & - & - & - \\

Number of zero-crossings & LibROSA & $(2.92\pm1.34)\times10^4$ & - & - & - & - & - & - & - & - & - & - \\

Loudness & pyloudnorm \cite{pyloudnorm} & $-24.5\pm2.89$ dB & $\downarrow$\cite{bayestehtashk2015fully} & $\downarrow$\cite{tomik2010dysarthria} & $\downarrow$\cite{noffs2018speech} & - & - & - & - & - & - & - \\

Loudness variation (sliding-window SD) & pyloudnorm & $5.80 \pm 2.72$ dB & $\downarrow$\cite{bayestehtashk2015fully} & $\downarrow$\cite{tomik2010dysarthria} & $\uparrow$\cite{noffs2018speech} & - & - & - & - & - & - & - \\

Crest factor*$\dagger$ & Ours & $4.35\pm1.15$ ($\mathrm{median}\pm\mathrm{MAD}$) & - & - & - & - & - & - & - & - & - & - \\ \hline

\textbf{Motivated by the clinical literature} && & & & & & & & & & & \\
\hline

Pitch period entropy* & Ours & $3.96 \pm 3.37$ ($\mathrm{median}\pm\mathrm{MAD}$) & $\uparrow$\cite{little2008suitability} & - & - & $\uparrow$\cite{rusz2014phonatory} & - & - & - & - & - & - \\

Jitter variants & - & - & $\updownarrow$\cite{bocklet2013} & $\uparrow$\cite{tomik2010dysarthria,silbergleit1997acoustic} & $\uparrow$\cite{noffs2018speech,dogan2007subjective,feijo2004acoustic} & $\uparrow$\cite{rusz2014phonatory,chan2019hd} & $\uparrow$\cite{low2020automated} & - & $\uparrow$\cite{low2020automated} & - & - & - \\
Jitter (local) & Ours & $0.0128 \pm 0.00374$ & - & - & - & - & - & - & - & - & - & - \\

Jitter (local, absolute) & Ours & $(9.31\pm2.97)\times10^{-5}$ s & - & - & - & - & - & - & - & - & - & - \\
Jitter (RAP) & Ours & $(3.14\pm0.928)\times10^{-3}$ & - & - & - & - & - & - & - & - & - & -  \\
Jitter (PPQ5) & Ours & $(5.53\pm1.62)\times10^{-3}$ & - & - & - & - & - & - & - & - & - & - \\
Jitter (DDP) & Ours & $(9.43\pm2.78)\times10^{-3}$ & - & - & - & - & - & - & - & - & - & - \\

Shimmer variants & - & - & $\uparrow$\cite{tsanas2012novel,bocklet2013} & - & $\uparrow$\cite{noffs2018speech,dogan2007subjective,feijo2004acoustic} & $\uparrow$\cite{rusz2014phonatory,chan2019hd} & $\uparrow$\cite{low2020automated} & - & $\uparrow$\cite{low2020automated} & - & - & - \\

Shimmer (local) & Ours & $0.0966 \pm 0.0231$ & - & - & - & - & - & - & - & - & - & - \\
Shimmer (local, db) & Ours & $0.737 \pm 0.113$ dB & - & - & - & - & - & - & - & - & - & - \\
Shimmer (APQ3) & Ours & $0.0363 \pm 0.00906$ & - & - & - & - & - & - & - & - & - & - \\
Shimmer (APQ5) & Ours & $0.0615 \pm 0.0161$ & - & - & - & - & - & - & - & - & - & - \\
Shimmer (APQ11) & Ours & $0.135 \pm 0.0497$ & - & - & - & - & - & - & - & - & - & - \\

Detrended fluctuation analysis & Ours & $0.940 \pm 0.152$ & $\uparrow$\cite{little2008suitability,tsanas2012novel,tsanas2011nonlinear} & - & - & $\uparrow$\cite{rusz2014phonatory,rusz2014phonatory} & - & - & - & - & - & - \\

Linear spectral coefficients & LibROSA & - & - & - & - & - & - & $\uparrow$\cite{low2020automated} & - & $\uparrow$\cite{low2020automated} & - & - \\

Linear spectral frequencies & Ours & - & - & - & - & - & - & - & - & - & - & - \\

Formant F1 & Ours & $(1.16\pm0.455)\times10^3$ Hz & $\updownarrow$\cite{rusz2013imprecise} & - & - & - & $\updownarrow$\cite{low2020automated} & $\uparrow$\cite{low2020automated} & $\updownarrow$\cite{low2020automated} & $\updownarrow$\cite{low2020automated} & - & - \\
Formant F2 & Ours  & $(1.93\pm0.468)\times10^3$ Hz & - & - & - & - & - & - & - & - & - & - \\
Formant F3 & Ours & $(2.73\pm0.452)\times10^3$ Hz & - & - & - & - & - & - & - & - & - & - \\
Formant F4 & Ours & $(3.52\pm0.453)\times10^3$ Hz & - & - & - & - & - & - & - & - & - & - \\

Formant $\Delta$F1$\dagger$ & Ours & $0.0417 \pm 1.17$  & - & - & - & - & - & - & - & - & - & - \\

Formant $\Delta$F2$\dagger$ & Ours & $0.0564 \pm 1.12$ & - & $\downarrow$\cite{tomik2010dysarthria} & $\downarrow$\cite{noffs2018speech} & - & - & - & - & - & - & - \\

Formant $\Delta$F3$\dagger$ & Ours & $0.0883 \pm 1.15$ & - & - & - & - & - & - & - & - & - & - \\

Formant $\Delta$F4$\dagger$ & Ours & $0.123 \pm 1.21$ & - & - & - & - & - & - & - & - & - & - \\

Sliding-window formant F1$\dagger$ & Ours & $(1.27\pm0.150)\times10^3$ Hz & - & - & - & - & - & - & - & - & - & - \\

Sliding-window formant F2$\dagger$ & Ours & $(2.20\pm0.140)\times10^3$ Hz & - & - & - & - & - & - & - & - & - & - \\

Sliding-window formant F3$\dagger$ & Ours & $(3.09\pm0.131)\times10^3$ Hz & - & - & - & - & - & - & - & - & - & - \\

Sliding-window formant F4$\dagger$ & Ours & $(3.96\pm0.137)\times10^3$ Hz & - & - & - & - & - & - & - & - & - & - \\

Sliding-window amplitude kurtosis$\dagger$ & Ours & $1.74 \pm 1.96$ & - & - & - & - & - & - & - & - & - & - \\

Amplitude shannon entropy & Ours & $(8.28\pm6.93)\times10^3$ & - & - & - & - & - & - & - & - & - & - \\

HNR & Ours & $9.11 \pm 2.29$ dB & $\downarrow$\cite{little2008suitability,tsanas2012novel} & - & - & $\downarrow$\cite{rusz2014phonatory,chan2019hd} & $\uparrow$\cite{low2020automated} & - & - & - & - & - \\ [1ex] 
\hline 
\end{tabular}}
\label{tab:table1} 
\end{table*}

\section{Surfboard Architecture} 
\label{section:lesson}

\subsection{Overview}
Surfboard aims to address the flaws of comparable frameworks while retaining their qualities. Specifically, we designed Surfboard with a focus on:
\begin{itemize}
    \item Ease of use within Python, the lingua franca of data scientists and ML engineers \cite{pythonrise1} and the primary language for ML frameworks such as PyTorch and TensorFlow.
    \item The ability to process large datasets, often needed for modern ML approaches to audio processing.
\end{itemize}
Like OpenSMILE, Surfboard extracts `components' (analogous to LLDs) as single values (e.g. loudness) or time series (e.g. MFCCs); in the latter case, statistics (e.g. the standard deviation) can be extracted from the time series to create time-independent features. These features can then be fed into e.g. a multilayer perceptron (MLP), a common use case for audio classification tasks. One can also obtain the full time series without extracting statistics for downstream sequential processing, for example using LSTMs \cite{hochreiter1997long}.

The audio features currently included in Surfboard are shown in Table \ref{tab:table1}. Inspired by some of the excellent work done by the audio community, Surfboard was built on top of a number packages that we found to be well-maintained, such as LibROSA \cite{mcfee2015librosa} and pysptk \cite{pysptk}. Further external implementations are referenced in Table \ref{tab:table1}. We first picked components and statistics which we deemed the most prominent in the ML audio/speech analysis literature, for example those in \cite{gemmeke2013exemplar, zhuang2010real, mesaros2010acoustic}. We then reviewed similar frameworks, including OpenSMILE and Praat, and conducted a clinical literature review of the application of speech to medical diagnosis to identify further features for inclusion. In \cite{tsanas2011nonlinear, tsanas2012novel, bocklet2013, teixeira2013vocal} for instance, the authors make use of features including the jitter variants, the shimmer variants and the harmonics-to-noise ratio to detect Parkinson's disease (PD) from speech. Another example from \cite{pramono2016cough} is the crest factor, which can be used for whooping cough detection. More details of motivation can be found in Table \ref{tab:table1}. 

Unlike OpenSMILE, Surfboard is released under an open-source license (the GNU GPL v3). We do this to ensure that the research community has the freedom to use and modify this software as they please, to empower new open source libraries built using Surfboard, and in the hope of fostering an active community of contributors.

\subsection{Using Surfboard}

There are two ways to use the package:
\begin{itemize}
    \item \textbf{Native Python:} The user imports the Surfboard module and instantiates the \texttt{Waveform} class to load an audio signal from an array or a file. Features are then extracted by calling \texttt{Waveform}'s methods, or in batches using the \texttt{extract\_features} helper function. This mode of use was designed for data exploration, medium-scale and on-the-fly feature extraction, for example when training a moderately sized ML model. The output is a Pandas \texttt{DataFrame} with each row representing the features extracted from a single waveform.
    \item \textbf{Command line interface (CLI):} The CLI is designed to extract features from a folder of audio files, given a configuration YAML file describing the desired features (i.e. the combination of components and statistics). The output is a CSV file corresponding to the \texttt{DataFrame} described above. See the documentation in our codebase for more details. We designed the CLI envisioning use for large-scale feature extraction requiring multiprocessing and/or submission to cloud virtual machine instances or local clusters. 
\end{itemize}

\subsection{Feature Comparison with OpenSMILE and Praat}

Surfboard, Praat and OpenSMILE extract different features, with a significant overlap. We consider a subset of the features (local shimmer, local jitter, DDP jitter) offered by all three frameworks. None of the three produces directly comparable values using default parameters due to substantial implementation differences. We therefore compare them using Spearman's rank correlation coefficient. The values obtained are given in Table \ref{tab:rhos}.

\begin{table}[ht]
\caption{Comparison of the three frameworks using Spearman's rank correlation coefficient. LJ = local jitter, DDPJ = DDP jitter, LS = local shimmer.} 
\centering 
\begin{tabular}{c c c c} 
\hline
Comparison & $\rho_{\text{LJ}}$ & $\rho_{\text{DDPJ}}$& $\rho_{\text{LS}} $ \\ [0.5ex] 
\hline 
 Praat vs OpenSMILE & 0.30 & 0.31 & 0.43\\ 
Surfboard vs OpenSMILE & 0.21 & 0.18 & 0.40\\
Surfboard vs Praat & 0.32 & 0.13 & 0.33\\ [1ex]
\hline 
\end{tabular}
\label{tab:rhos} 
\end{table}
We note significant rank correlation between the chosen features across all three pairs of frameworks. However, none of the three pairs agree perfectly. This can be attributed to differing implementations; for instance, Surfboard makes use of the RAPT pitch-tracking algorithm \cite{talkin1995robust} inspired by \cite{tsanas2011nonlinear}, whereas OpenSMILE and Praat each employ custom peak-picking algorithms (see \cite{boersma2018praat} and the OpenSMILE codebase\footnote{\url{https://www.audeering.com/opensmile/}}) for jitter/shimmer calculations. Furthermore, feature extraction functions typically require parameters (e.g. sampling rate of the waveform and hop length); here we choose framework-dependent default parameters, which could impact the rank correlations.

\section{Application: Classifying Parkinson's Disease Using the mPower Dataset} 

\subsection{Experimental Design}

A substantial number of the components developed in the Surfboard package were motivated by the clinical literature. Speech changes have been reported in a multitude of diseases (see Table \ref{tab:table1}); in this section we take the example of Parkinson's disease (PD). The main symptoms of PD include tremor and rigidity but effects on the motor system extend to the vocal cord, where vocal impairments are common \cite{hughes1993clinicopathologic,hanson1984cinegraphic}, with up to 70-90\% prevalence after the onset of the disease \cite{ho1999speech,logemann1978frequency}. These vocal impairments may be one of the earliest indicators of disease \cite{harel2004variability} and deterioration of speech accompanies PD progression \cite{harel2004variability}. These early signs include reduced voice volume (hypophonia) and breathiness, hoarseness or creakiness in the voice (dysphonia), preceding more generalized speech disorder \cite{ho1999speech,logemann1978frequency}. Such impairments can be detected using audio analysis; in \cite{little2008suitability}, the authors collected sustained phonations from 54 participants and selected 10 measures to use as input to a classifier, achieving 91.4\% accuracy. A follow-up study extended this set of features and achieved 99\% accuracy \cite{tsanas2012novel}. However, \cite{orozco2016automatic} criticized their work by arguing that the dataset was small (263 phonations) and that their training and test sets featured the same participants, an antipattern in ML that can lead to scientifically invalid conclusions.

We illustrate the application of Surfboard to the task of classifying PD sufferers versus healthy controls (HC). We work with the mPower dataset, a large, real-world dataset of sustained phonations from PD sufferers and HCs. Illustrating how test set design can be leveraged to achieve more rigorous evaluation metrics, we first create a subset $S$ of 12,094 of the 62,609 phonations, leaving 50,515 for training. We ensure that $S$ is balanced in terms of diagnosis (6,157 labelled PD and 6,157 labelled HC) and further split $S$ into three test sets, $S_1$, $S_2$ and $S_3$, such that:
\begin{itemize}
    \item $S_1$ comprises 2,500 phonations labelled HC and 2,500 phonations labelled PD, randomly sampled from $S$. We do not restrict the inclusion of phonations from participants already present in the training set.
    \item $S_2$ comprises 2,547 phonations labelled HC and 2,547 phonations labelled PD, randomly sampled from $S \setminus S_1$ with the additional constraint that the phonations contained in $S_2$ cannot be produced by participants already in the training set (i.e. $S_2$ and the training set are disjoint by participant).
    \item $S_3$ comprises the remaining 1,000 HC participants and 1,000 PD participants. We ensure that $S_3$ is age- and gender-matched, and that $S_3$ and the training set are disjoint by participant.
\end{itemize}

We extract features using Surfboard v0.1, choosing the subset of Surfboard components with demonstrated clinical relevance to PD (see Table \ref{tab:table1}): MFCCs, jitters, shimmers, PPE, HNR, loudness, formants, log energy, RMS. We use the entire statistics set offered by Surfboard to generate feature vectors from the time series components, resulting in one 377-dimensional vector per phonation. If feature extraction fails, for example if the F0 extraction fails to recognize voiced segments for a given phonation, we replace missing fields with column averages, as is common with tabular data (albeit flawed \cite{baraldi2010introduction}).

We use the remaining 50,515 phonations to train a single gradient boosting classifier using the scikit-learn library and evaluate separately on $S_1$, $S_2$ and $S_3$. We hypothesize that a decrease would be seen in our trained classifier's performance between $S_1$ and $S_2$, since the classifier can no longer benefit from merely identifying participants. We likewise expect a deterioration in performance between $S_2$ and $S_3$, since the classifier can no longer learn to leverage the difference in age and gender distributions between the PD and HC subsets of the data.

\subsection{Results}

The results are shown in Table \ref{tab:classification}, illustrating a progressive decrease in performance. This supports our hypothesis and the criticism raised by \cite{orozco2016automatic}. It is worth noting that the largest decrease in performance comes between $S_2$ and $S_3$, the unmatched and matched subsets respectively. In this work, we do not carry out a thorough feature selection process to optimize classification metrics. Instead, we illustrate how Surfboard can be used for audio analysis and exhibit some of the prior flaws in the existing literature, hopefully stimulating additional statistical rigor in future PD research.

\begin{table}[ht]
\caption{PD/HC classification results on the three test sets. See main text for a description of $S_1$, $S_2$ and $S_3$.} 
\centering 
\begin{tabular}{c c c c c} 
\hline
Test set & Precision & Recall & Accuracy & AUC \\ [0.5ex] 
\hline 
$S_1$ & 0.77 & 0.74 & 0.74 & 0.85 \\ 
$S_2$ & 0.75 & 0.72 & 0.72 & 0.79 \\
$S_3$ & 0.64 & 0.63 & 0.63 & 0.69 \\ [1ex]
\hline 
\end{tabular}
\label{tab:classification} 
\end{table}


\section{Conclusion}

This paper presented Surfboard, a Python package for clinical audio analysis designed for modern ML. We described its high-level architecture and the rationale behind our design choices; we borrowed what we deemed relevant from existing frameworks and discarded flaws in the context of ML workflows. We compared a subset of features extracted from Praat, OpenSMILE and Surfboard and built a toy example on voice data from mPower, demonstrating the clinical relevance of Surfboard features. Our hope is that Surfboard will enable researchers to extract useful features from large datasets, enabling faster and more consistent research in this space. We also hope that it will inspire novel research in the field of audio analysis, particularly at its intersection with the clinical domain.

\section{Acknowledgements}

We would like to acknowledge and thank the participants of the mPower study who made our use case of classifying PD from sustained phonations possible. This data was contributed by users of the Parkinson's mPower mobile application as part of the mPower study. The study was run by Sage Bionetworks and data access is managed via the Synapse platform\footnote{\url{https://www.synapse.org/}}.

\bibliographystyle{IEEEtran}
\bibliography{mybib}
\end{document}